\pdfoutput=1

\documentclass[sigconf, nonacm, screen]{acmart}

\usepackage{algorithmic}
\usepackage{graphicx}
\usepackage{textcomp}
\usepackage{xcolor}
\usepackage{soul}
\usepackage[braket, qm]{qcircuit}
\usepackage[edges]{forest}
\usetikzlibrary{arrows.meta}

\begin{document}

\date{}

\title{Classification of Quantum Computer Fault Injection Attacks}

\author{Chuanqi Xu}
\affiliation{%
  \institution{Yale University}
  \city{New Haven}
  \state{CT}
  \country{USA}
}
\email{chuanqi.xu@yale.edu}

\author{Ferhat Erata}
\affiliation{%
  \institution{Yale University}
  \city{New Haven}
  \state{CT}
  \country{USA}
}
\email{ferhat.erata@yale.edu}

\author{Jakub Szefer}
\affiliation{%
  \institution{Yale University}
  \city{New Haven}
  \state{CT}
  \country{USA}
}
\email{jakub.szefer@yale.edu}

\begin{abstract}
The rapid growth of interest in quantum computing has brought about the need to secure these powerful machines against a range of physical attacks. As qubit counts increase and quantum computers achieve higher levels of fidelity, their potential to execute novel algorithms and generate sensitive intellectual property becomes more promising. However, there is a significant gap in our understanding of the vulnerabilities these computers face in terms of security and privacy attacks. Among the potential threats are physical attacks, including those orchestrated by malicious insiders within data centers where the quantum computers are located, which could compromise the integrity of computations and resulting data. This paper presents an exploration of fault-injection attacks as one class of physical attacks on quantum computers. This work first introduces a classification of fault-injection attacks and strategies, including the domain of fault-injection attacks, the fault targets, and fault manifestations in quantum computers. The resulting classification highlights the potential threats that exist. By shedding light on the vulnerabilities of quantum computers to fault-injection attacks, this work contributes to the development of robust security measures for this emerging technology.
\end{abstract}

\maketitle

\section{Introduction}

Quantum computing has accelerated in development in recent years. Many companies and universities are racing to build bigger and better machines. Among others, IBM unveiled a 433 qubit quantum computer in late 2022. A 4158 qubit IBM quantum computer is projected for 2025, and most recently IBM has announced plans for 100,000 qubit quantum computers by 2033~\cite{ibmChartingCourse}.

Presently, quantum computers are in the Nosy Intermediate Scale Quantum (NISQ) regime~\cite{Preskill2018quantumcomputingin}, where there are only quantum computers with qubits less than 1000 and are not able to support quantum error correction~\cite{Devitt_2013} and practical applications like Shor's algorithm~\cite{doi:10.1137/S0097539795293172}. Nevertheless, these machines have the potential to help accelerate drug discovery or find new materials. As the machines grow in size and fidelity they may reach quantum supremacy, which refers to a point in time at which a quantum computer can perform calculations that significantly surpass the capabilities of even the most powerful classical computers. Already, some studies have shown quantum supremacy before fault-tolerant~\cite{arute2019quantum, kim2023evidence}, though such an assertion may be controversial at this point. With the increase of qubits and improvement in fidelity, it will be possible to gradually move into the fault-tolerant quantum computing regime with techniques like quantum error correction. Optimistically, quantum computers and quantum algorithms promise to be applied to revolutionize many fields, such as Grover's~\cite{10.1145/237814.237866} and Shor's algorithms that can be used to break some nowadays widely-used classical cryptographic algorithms like RSA~\cite{10.1145/359340.359342}.

As quantum computers grow in size, the data and information in the computing process may be sensitive and private. Further, the quantum programs themselves executed on quantum computers are also valuable intellectual properties. Integrity and confidentiality of the data or quantum programs can be compromised if there is a fault-injection attack.

\subsection{Differences from Classical Computer Fault Injection}

In classical computer fault injection, the faults mainly target the instructions executing on the processor or the data in registers. It is also possible to inject or cause faults in DRAM memory or on the memory bus or other parts of the system. The classical processor is typically encased in a single package, and in fault-injection attacks, the package is exposed to voltage glitching, clock glitching, EM, lasers, or other sources of disturbance. Section~\ref{sec:related_work} provides a brief list of existing works on fault injection in classical computers.

One main difference in quantum computers is the extensive classical infrastructure that controls the qubits within the quantum computers. This infrastructure significantly extends the possible attack surface. Also, given the current and projected physical size of quantum computers (the size of a server room or at least a few server racks), physical access and opportunity to manipulate the equipment is much larger than with today's nanometer-sized transistors in CPUs. Further, there is an opportunity for attackers to either manipulate the qubits, or classical registers into which the qubit measurements are read, or the control signals (either digital signals going into the controller equipment, or analog signals going between controller equipment and the quantum computer itself). This extended attack surface compared to classical computers analyzed in this work sheds light on the possibly new perspective in quantum computer fault injection attacks.

\subsection{Contributions}

The contributions of this work are:

\begin{itemize}
    \item We identify the {\em domain of quantum computer fault injection attacks}; this domain represents the attack surface that is distinct from classical computers, and at the same time it identifies the hardware components that may be subject to the fault injection attacks.
    \item We pinpoint 3 {\em fault targets} specific to quantum computers: quantum processing units, quantum computer controller, and classical co-processors; within the three targets, we present further 6 specific components that can be targeted for fault attacks.
    \item We present {\em fault model}, {\em fault bound}, and {\em fault lifespan} for the different fault targets.
    \item We propose the first classification of quantum computer fault injection attacks to help industry and researchers navigate the security of this emerging technology.
\end{itemize}

\begin{figure*}[t]
  \centering
  \includegraphics[width=0.8\textwidth]{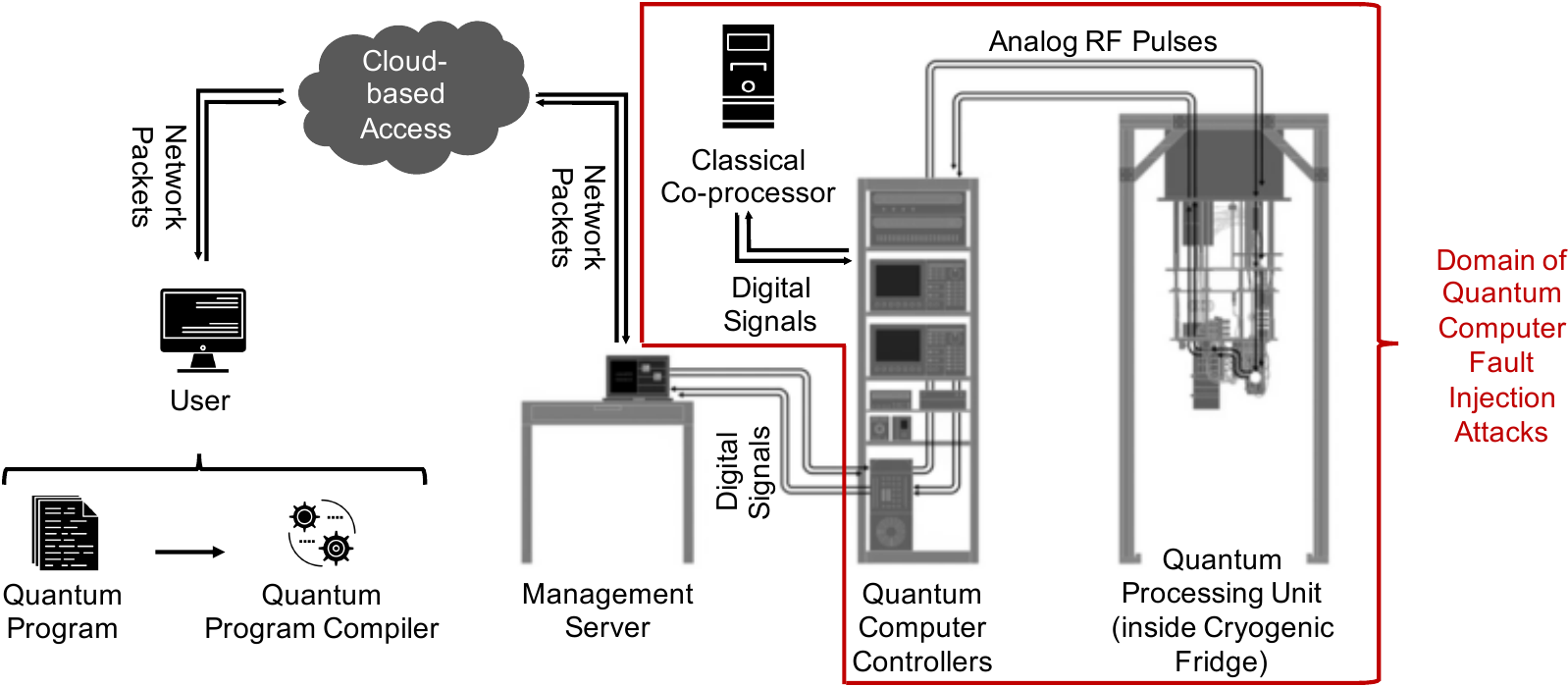}
  \caption{\small Typical setup of a  superconducting qubit quantum computer, figure is based on figure provided by IBM.}
  \label{fig_attack_surface}
\end{figure*}

\section{Background}
\label{sec_background}

This work focuses on superconducting quantum computers, such as those available from IBM, Rigetti, QCI, and others. The typical setup of a  superconducting qubit quantum computer is shown in Figure~\ref{fig_attack_surface}. We consider today's cloud-based computers where users connect remotely to the machines. Figure~\ref{fig_attack_surface} specifically depicts a superconducting qubit quantum computer setup. Other types of quantum computers may have different types of, for example, quantum computer controllers, but the same types of fault injection attacks can be applied.

\subsection{Quantum Computing Basics}

Analogous to the classical bit, a quantum bit, or \textit{qubit}, is the fundamental computational unit in quantum computers. A qubit can be represented with the bra-ket representation. With $\ket 0$ and $\ket 1$ as the basis states, a qubit can be written as $\ket \psi = \alpha \ket 0+ \beta \ket 1$, where $|\alpha|^2 + |\beta|^2 = 1$. According to Born’s rule, the results of measuring $\ket \psi$ is either $\ket 0$ or $\ket 1$, with probability $|\alpha|^2$ and $|\beta|^2$ respectively. Such a phenomenon that a qubit can be measured with two results is not seen in classical computing, and it is often called \textit{superposition}. Also, the state after the measurement will \textit{collapse} to the resulting state, no matter what the initial state is. Similarly, an $n$-qubit system is spanned by $2^n$ basis states. Surprisingly, some multi-qubit quantum states cannot be described independently by the state of their components, which is another phenomenon that is not shown in classical computing, and this is often referred to \textit{entanglement}. 
Qubits are controlled and evolved by \textit{quantum gates}, which are the building blocks of quantum circuits, like classical logic gates are for conventional digital circuits. We refer interested readers to~\cite{nielsen2001quantum} for details.

\subsection{Cloud-based Access}

Due to the expensive nature of quantum computing equipment, quantum computers are currently available as cloud-based systems. For example, cloud-based services such as IBM Quantum~\cite{ibmquantum}, Amazon Braket~\cite{amazonbracket}, and Azure Quantum~\cite{azurequantum} already provide access to Noisy Intermediate-Scale Quantum (NISQ) quantum computers remotely for users. In the cloud setting, the user has no control over the management server, quantum computer controllers, and the cryogenic fridge are not under the control of the user. A malicious insider or compromised cloud provider could try to perform fault injection attacks.

\begin{figure*}[t]
  \centering
  \includegraphics[width=0.99\textwidth]{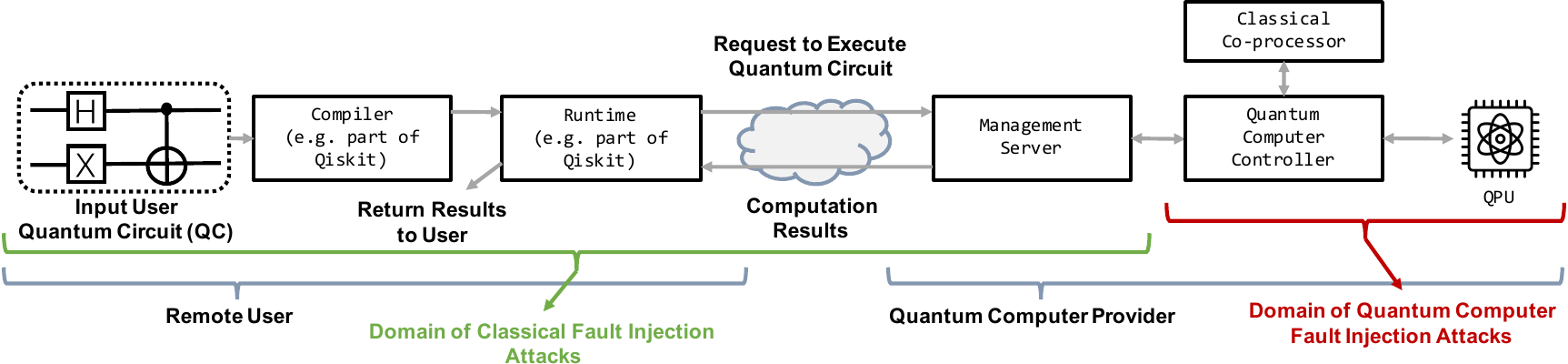}
  \caption{\small Typical quantum computer workflow.}
  \label{fig_workflow}
\end{figure*}

\subsection{Users' Quantum Program}

Quantum programs may be described using `gates'. When using gate-level description, programs are composed of quantum gates. Quantum computers typically do not and it is also not necessary for them to support all kinds of quantum gates, because it is proved that any quantum gate can be approximated within a minor error using only a small number of quantum gates~\cite{doi:10.1098/rspa.1995.0065}. As a result, before running quantum circuits on a target computer, they are required to be transformed into quantum circuits that only contain the supported quantum gates, usually called \textit{native gates} or \textit{basis gates}, in order to be compatible with target quantum computers. This process is often called \textit{transpilation} or \textit{compilation}. Besides to fulfill the native gates requirement, transpilers or compilers that do this step also need to satisfy other restrictions, such as that qubit connections need to be available on target quantum computers. Other processes such as optimizations may also be involved in~transpilation. 

In the end, quantum circuits only including native gates need to be further transformed into corresponding low-level operations and signals. This is a lower-level description of quantum circuits. For superconducting quantum computers, qubits are controlled by analog radiofrequency (RF) pulses, which are sent to the quantum computer. When using this \textit{pulse-level} description, users directly specify the control pulses used to trigger the quantum gate operations. The pulses are very specific to each machine and typically need to be carefully designed to perform the desired operation. Advanced users and researchers may choose to use the pulse-level approach, rather than using the pre-defined native gates.

\subsection{Management Server}

The management server is a typical classical server that sits between the users and the quantum equipment. Management servers in the context of cloud computing oversee and control resources hosted on a cloud platform. 
Management servers for quantum computing commonly handle the receiving of quantum jobs, queuing, and dispatching jobs. Quantum jobs submitted by users are usually first pushed into priority queues, and based on the priority algorithms of the cloud platforms, these jobs wait in the queue, and then the information of jobs is processed and sent to quantum computer controllers after they finish waiting.

\subsection{Classical Co-processor}

A classical co-processor is a classical computer part of the quantum computer controller, or tightly coupled to the controller. The co-processor can perform classical computations based on the data readout from the quantum computer. It may contain user-defined code or application-specific code defining what operations to perform based on the readout data; as well as it can be used to determine what subsequent operations to execute on the quantum computer or to update the circuit executing on the quantum computer. In one example of quantum machine learning (QML)~\cite{biamonte2017quantum}, based on the readout data, the co-processor can optimize the parameters of the quantum circuit and issue the next job with the updated circuit, similar to the classical machine learning.

\subsection{Quantum Computer Controller}
\label{sec:quantum_controllers}

In current small-scale quantum processors, each qubit or qubit pair is typically assigned dedicated control pulses with distinct parameter settings, including the pulse waveform, pulse duration, pulse frequency, pulse amplitude, and so on. Control pulses, both microwave and baseband flux, are generated at room temperature by classical equipment such as the arbitrary waveform generator (AWG) and IQ mixers. Then these pulses will be delivered to the qubits in the cryogenic system through a series of attenuators and filters designed to suppress harmful noises when the quantum programs reach the point to run the corresponding gates. 

Besides controlling the qubits, one important function of quantum computer controllers is to perform the measurement process and measurement readout results. The results from quantum computers may be stored in the controller and sent back to the management servers when jobs finish. In addition, for advanced features like dynamic circuits~\cite{ibmFullPower}, it stores the middle-measurement results and controls future operations based on these results.

\subsection{Quantum Processing Unit including Cryogenic Fridges}

The Quantum Processing Unit (QPU) contains the actual physical qubits. The QPU is located in the cryogenic fridge, also known as the dilution refrigerator, which is an integral part of superconducting quantum computers. 
These qubits are sensitive to thermal noise, which is why the frigid environment provided by the dilution refrigerator is crucial. Once the qubits are in their superconducting state, they are manipulated using microwave pulses, generated by quantum computer controllers previously introduced. The pulses are delivered through coaxial cables that are also cooled within the refrigerator to minimize thermal noise. 

In the end, the qubit states are usually measured to get the job results. In superconducting quantum computers, measurement equipment often includes cryogenic amplifiers and analog-to-digital converters. These tools discern the quantum state of the qubits by monitoring microwave signals. The qubit-induced changes in these signals are amplified and converted into digital information, making quantum data readable to classical computers. Then this data is sent back to the quantum computer controllers, which then send it to the management server; and the users finally are forwarded the measurement data.

\subsection{Workflow of Executing Quantum Circuits on a Quantum Computer}

The typical workflow of quantum computers is shown in Figure~\ref{fig_workflow}. In quantum computing, users can write gate-level programs using quantum programming languages such as Qiskit~\cite{Qiskit}, Amazon Braket SDK~\cite{braket_sdk}, or Cirq~\cite{cirq_developers_2022}. 
These programs consist of sequences of quantum gates that operate on qubits. The programs are then transpiled to decompose the gates into elementary quantum gates supported by the hardware. The transpiler optimizes the program by reducing gate count and improving gate ordering. It also maps logical qubits to the physical qubits available in the hardware, considering connectivity constraints. The next step is \textit{scheduling}, where timing and control information are determined for each gate, specifying the precise microwave pulses required for their execution. When jobs are sent to quantum computer systems and start to execute, microwave electronics generate these pulses, corresponding to signals that manipulate the quantum state of the qubits. The pulses are applied to the physical qubits, implementing the desired gate operations. After execution, the resulting quantum state can be measured to obtain the computation's output. The specific details of the transpilation and scheduling process may vary depending on the programming language, hardware, and software stack used.

\section{Fault Manifestation}

In the context of fault injection in quantum computing, \textit{fault manifestation} refers to the observable effect or consequence of an injected fault within the quantum system. This could include changes in the state of a qubit, alterations in the operation of a quantum gate, or eventually deviations in the outcome of a quantum algorithm. The study of fault manifestation is crucial in understanding the impact of errors on quantum computations and in developing strategies for error detection and correction. The fault manifestation can be:

\subsection{Gate-level Program}

The gate-level quantum circuit is a model used in quantum computing to describe qubit evolution and incidental operations. Computations in the gate-level quantum circuits are represented as a sequence of quantum gates acting on qubits, and other operations such as measurement, reset, and classical operations, from left to right to denote time steps. Each quantum gate, analogous to a logic gate in classical computing, performs a specific unitary operation or transformation on the quantum state of a qubit or a set of qubits. By arranging these gates in specific sequences and combinations, complex quantum algorithms can be implemented. This gate-level description is particularly useful for visualizing, designing, and analyzing quantum computations. 

\begin{figure}[t]
  \centering
    \includegraphics[width=\columnwidth]{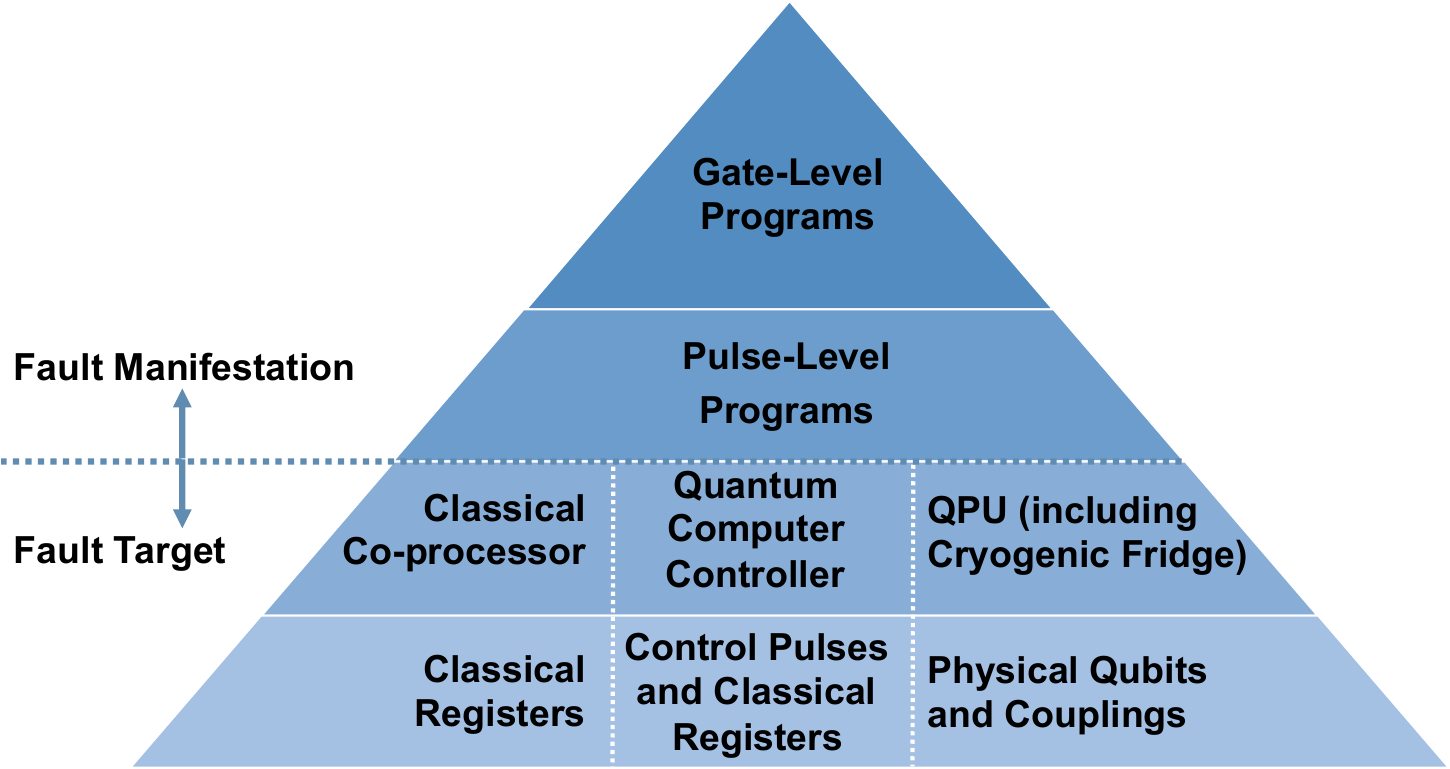}
  \caption{\small The fault target and fault manifestation security pyramid for superconducting quantum computers.}
  \label{fig_fault_targets}
\end{figure}

\subsection{Pulse-level Program}

This is one level lower abstraction of quantum circuits. Since superconducting qubits are controlled by microwave pulses, the exact physical actions of quantum gates and other operations in gate-level circuits are correspondingly predefined microwave pulses. The pulse parameters such as frequency and amplitude are continuously changing due to the fluctuations in the environment and qubits. Therefore, the pulse parameters are frequently calibrated to reach high fidelity to the desired logic operations specified by the corresponding quantum gates. A pulse-level description provides a more granular view of quantum computation compared to the gate-level representation. It accounts for the physical implementation of quantum gates, offering insights into the precise control mechanisms and potential sources of error in quantum operations.

\section{Fault Target}
\label{sec_fault_targets}

Faults can occur or be injected at various locations or types of equipment within the quantum computing system. We focus here on the components within the domain of quantum computer fault injection attacks, defined in Figure~\ref{fig_attack_surface}: Quantum Processing Unit, Quantum Computer Controller, and Classical Co-processor.

\subsection{Classical Co-processor}
\label{sec_fault_target_coproc}

The classical co-processor is a classical equipment made of traditional digital computing hardware and peripherals, which is introduced in Section~\ref{sec:quantum_controllers}. The classical co-processor may be used in conjunction with the quantum computer. For example, in quantum machine learning (QML), there is an iterative process of running a circuit on a quantum computer, optimizing the circuit on a classical computer based on results, running it again on a quantum computer with updated parameters, etc.

\subsubsection{Faults in Classical Registers}

Within the classical co-processor are of course the usual components such as ALU, registers, or memory, among others.%
\footnote{For simplicity, we specify the fault target here as ``classical registers'', but the physical faults could also be in ALU, memory, or other components. Since the faults will eventually occur in or enter registers, we use the simplification of calling the target just ``classical registers''.}
Faults can be injected in these classical components to, for example, affect the computations used in QML optimization routines between executions of a circuit on a quantum computer. For program specification at the gate-level, the faults can result in gates being added, removed, or modified by changing the digital bits that specify them in the program. For program specification at the pulse-level, the faults can affect the digital specification of the amplitude, duration, or phase of the control pulses to be~generated.

\subsection{Quantum Computer Controller}
\label{sec_fault_targets_controller}

Quantum computer controller is typically made of equipment to generate microwave pulses to manipulate qubit states, and measurement equipment to translate quantum information into a classical format which is stored as the readout data.

\subsubsection{Faults in Control Pulses}

Faults can be injected into the control pulses generated by the quantum computer controller, for example, through EM radiation that affects the pulses generated by the controller, or more directly by affecting the operation of the controller itself causing it to generate wrong or modified pulses. Readout data is the classical data resulting from the measurements. Faults can also be injected into the readout control pulses through EM, for example, or the readout data can itself be directly manipulated through faults in digital registers storing the data within the~controller.

Faults can occur or be induced during the application of the control pulses that manipulate the quantum state of the qubits. In particular, faults can arise or be induced in the control electronics that generate and apply microwave pulses to manipulate the qubits. Issues with signal generation, calibration, timing, or stability of control signals can impact the accuracy of gate operations. Faults can affect unitary and non-unitary operations:

\begin{itemize}

    \item \textit{Unitary Operations} -- Unitary operations refer to transformations that preserve the normalization and reversibility of quantum states. Quantum gates are unitary gates, and unitary operations are the typical computational operations on the qubits, such as different X, SX, CX, or other gates.

    \item \textit{Non-Unitary Operations} -- Non-unitary operations are all other operations. For instance, reset or measurement are not unitary, because they collapse the state of the qubits during the execution of the operation.
    
\end{itemize}

\subsubsection{Faults in Classical Registers}

Non-unitary operations such as reset or measurement utilize classical registers. In particular, when qubits are measured, the quantum state collapses to one of the eigenstates of the measurement, and the measurement result is stored in classical registers or memories inside the control electronics. The classical registers then can be victims of fault injection that affects the classical bits. These faults can affect operations in which the classical registers participate, such as mid-circuit measurement, and final measurement:

\begin{itemize}

    \item \textit{Mid-Circuit Measurement} -- Mid-circuit measurement allows for measuring the qubit state in the middle of the execution. The results can then be used to determine what code to execute by analyzing the classical bit measurement results. If the classical bit is modified, the circuit execution can be affected, as the classical bit at each mid-circuit measurement determines the next set of operations that will be applied. One example is the reset instruction, where a measurement is performed to measure the qubit state, and if the state is measured to be 1 instead of 0, an $X$ gate is applied to flip the qubit back to $\ket 0$. If the classical register is modified, then the reset will reset to the other state.

    \item \textit{Final Measurement} -- The final measurement is performed at the end of each circuit. Usually, all qubits are measured, though sometimes ancilla qubits may not be measured. Injecting fault into the classical bits at this stage is effectively equivalent to manipulating the final circuit output.

\end{itemize}

\subsection{Quantum Processing Unit}

The quantum processing unit (QPU) refers to the hardware components in a superconducting quantum computer that operate based on the principles of quantum mechanics. This includes the QPU which implements qubits, such as the Josephson junction widely used to realize superconducting qubits.

\subsubsection{Faults in Physical Qubits or Couplings}

There are many ways to influence and thus inject faults into the qubits. For instance, superconducting qubits are susceptible to decoherence, which refers to the loss of coherence and information due to interactions with the environment. External noise sources, such as thermal fluctuations or electromagnetic radiation, can cause qubits to lose their quantum states and result in errors. Faults can be injected through external means such as EM radiation or thermal changes to the fridge holding the qubits.

\section{Classification}

\begin{figure*}[t]
    \centering
    \includegraphics[width=\textwidth]{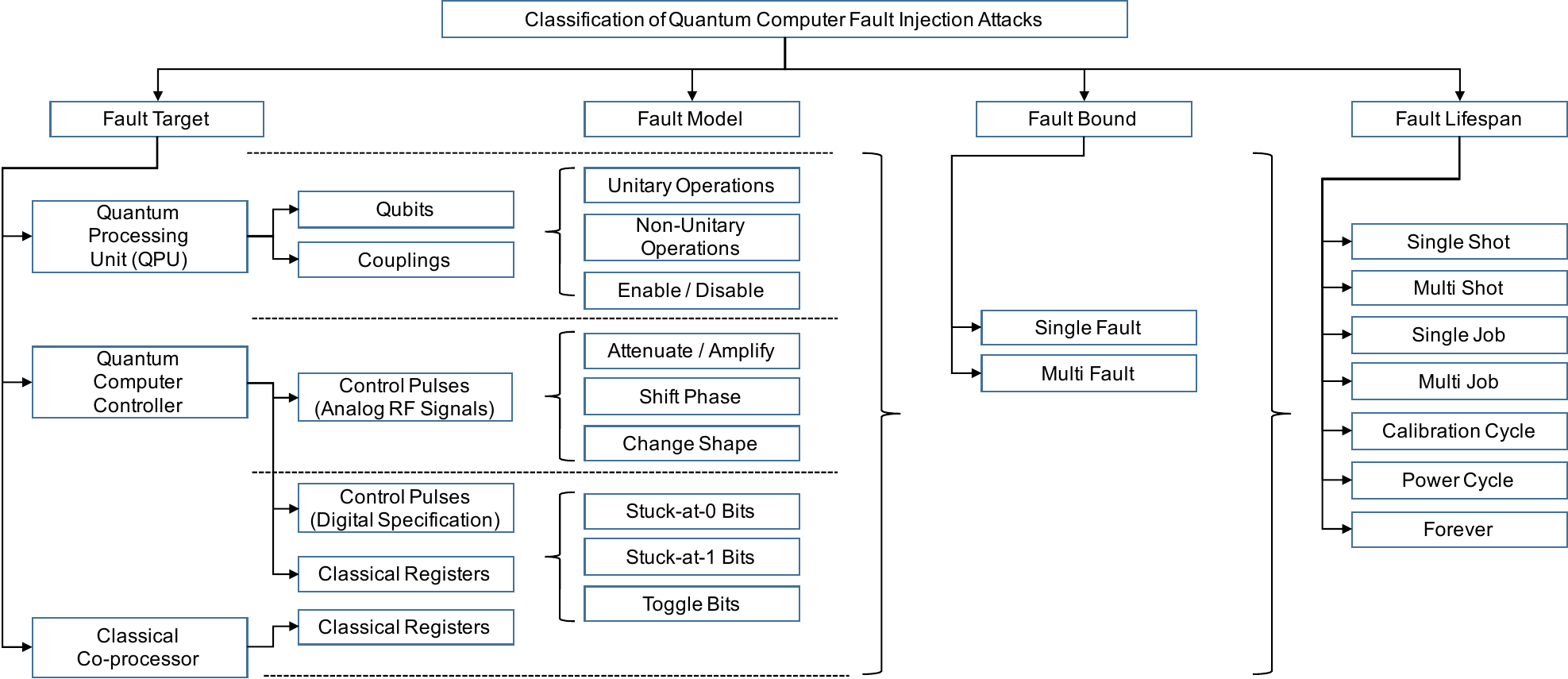}
    \caption{\small Classification of quantum computer fault injection attacks.}
    \label{fig_classification}
\end{figure*}

Our classification of quantum computer fault injection
attacks is now presented in this section. 
The classification is presented in Figure~\ref{fig_classification} and detailed below.%
\footnote{The terminology used in this section focuses on superconducting qubit machines, but this classification can be equally applied to other types of quantum computers by replacing certain terms. For example, control microwave pulses can be replaced by laser pulses if ion-trap computers are considered.}

\subsection{Fault Targets}

In the classification, we separate the three targets into six specific components vulnerable to faults and list them in more detail below.\\

\textbf{Quantum Processing Unit:}
\begin{itemize}
        \item Target: \textit{Qubits} are typically physical, two-level quantum-mechanical systems. A common type of qubit is built from a Josephson junction (but many others exist). As physical systems, they can be impacted by voltage changes, EM radiation, etc., that attackers can generate.
        \item Target: \textit{Couplings} are typically intermediate electrical circuits used to connect qubits, they can be likewise impacted by voltage changes, EM radiation, etc., that attackers can~generate.
    \end{itemize}

\textbf{Quantum Computer Controller:}
    \begin{itemize}
        \item Target: \textit{Control Pules (Analog RF Signals)} are often microwave pulses sent to an antenna or transmission line coupled to the qubit with a frequency resonant with that qubit to realize an operation. The attacker can interfere with or modify analog properties of the signals to induce faults in the qubits or gate operations, e.g., by changing the frequency, phase, or envelope.
        \item Target: \textit{Control Pulses (Digital Specification)} are generated by arbitrary waveform generators from digital specification, e.g. by an FPGA. The attacker performs attacks on classical bits or classical operations that read, modify, or write the digital information, thus resulting in wrong pulses being sent.
        \item Target: \textit{Classical Registers} are used, for example, to store measurement readout information during mid-circuit or final measurement. In particular, the mid-circuit measurement may be used to determine subsequent operations in dynamic circuits~\cite{ibmFullPower}. The attacker can induce faults in these classical ~registers.
    \end{itemize}
    
\textbf{Classical Co-processor:}
    \begin{itemize}
        \item Target: \textit{Classical Registers} are also used in classical co-processors used to perform computations on the output. For example in quantum machine learning (QML), parts of the input circuit are optimized based on the results of computation, and the circuit is run again. The attacker can induce faults in these classical registers. 
    \end{itemize}

\subsection{Fault Model}

The fault model is a theoretical representation or framework that predicts or describes the types of faults that may occur in a system, their causes, and their potential effects. We have three fault models, corresponding to different targets.

\textbf{Quantum Processing Unit:} The qubits and couplings are vulnerable to three types of novel faults not found in classical computers: Faults can result in unitary type operations, which are effectively faults inducing a change in qubit state that can be reversed like any other (non-malicious) unitary gate. Faults can result in non-unitary operations, which are usually hard to reverse. Faults can result in enabling / disabling of qubits or couplings, which may be similar to instruction skip faults in classical computers if a coupling is disabled, for example.

\textbf{Quantum Computer Controller:} The analog control pulses are also vulnerable to novel types of faults not found in classical computers: Faults can attenuate / amplify the analog pulses, causing different gate operations to be effectively performed. Faults can also shift the phase of the pulses, likewise resulting in different gate operations being effectively performed. The faults can also change the shape of the envelope of the pulse, again changing the gate operation performed. If the pulses are attenuated or otherwise sufficiently distorted, a gate operation may be effectively disabled. Conversely, amplifying or otherwise injecting an analog signal can create or insert a gate operation not part of the original circuit.

\textbf{Quantum Computer Controller and Classical Co-processor:} The controller and the co-processor also contain digital classical information, specifying the pulses (before they are generated as analog microwave signals) and other registers. These are vulnerable to well-known stuck-at faults or bit toggling faults.

\subsection{Fault Bound}

The fault bound is a limit or threshold that defines the maximum number of faults that a system can tolerate without significant degradation in its performance or functionality. Regardless of the fault target, there is either a single or multiple fault threat.

\subsection{Fault Lifespan}

The fault lifespan refers to the duration for which a fault persists in a system. For quantum systems, this could refer to the period during which a qubit remains in an erroneous state before it is corrected or resets to its original state. In quantum computers, there are many more different lifespans compared to classical computers.

\begin{itemize}
    \item Single Shot -- each circuit is divided into one or more shots that are executed on a quantum computer; most short-lived faults would affect single shots. Most faults on analog pulses would fit in this category.
    \item Multi Shot -- faults can persist through the execution of multiple shots of a circuit. Modification of the digital specification of the pulses would fit in this category.
    \item Single Job -- multi-shot faults that last for all shots of a circuit would be Single Job faults.
    \item Multi Job -- faults across multiple jobs of the same or different users would be multi-job faults. Faults in classical co-processor registers could fit in this category.
    \item Calibration Cycle -- each quantum computer is calibrated frequently. Calibration can correct for changes in the environment or noise. Unitary operation-type faults in qubits could be in this category.
    \item Power Cycle -- periodically, a quantum computer fridge has to be warmed up to replace or modify hardware, this is effectively a power cycle. Changes to control pulses which cause rapid heating and then cooling of the qubits could result in flux trapping, requiring power cycling the fridge.
    \item Forever -- faults that permanently alter the hardware would be faults that last forever. Disable faults on couplings could fit in this category.
\end{itemize}

\section{Related Work}
\label{sec:related_work}

There are only a few studies on fault injection attacks in quantum computers. Most of them are based on the hardware-induced faults in qubits~\cite{9833778, vepsalainen2020impact, google2021exponential}. Therefore, we drew inspiration from the fault injection literature in classical computing. Our decomposition includes Fault Target, Fault Model, Fault Bound, and Lifespan~\cite{baksi2022survey, verbauwhede2011fault, shepherd2021physical}. However, our classification represents the attack surface that is distinct from classical computers and, at the same time, identifies the hardware components that may be subject to fault injection attacks in quantum computers.

Giraud et al.~\cite{giraud2004survey} classify fault injection attacks in classical computing as transient vs. permanent and invasive vs. non-invasive. However, for our study, we focused solely on non-invasive attacks and classified them as transient or permanent under the Fault Lifespan category.
In a recent study by Ravi et al.~\cite{ravi2022side}, fault injection attacks specific to post-quantum cryptography algorithms (Kyber and Dilithium) were classified. This attack-specific classification examined characteristics such as the need for profiling, the number of required traces, and the ability to observe or communicate with the victim device.
Baksi et al.~\cite{baksi2022survey} conducted a survey on fault attacks on symmetric key cryptosystems, consolidating existing attacks under fault models, data alteration methods, sources of fault injection, and analysis methods. Although our fault injection attack classification for quantum computers shares similar categories, it is tailored specifically to the quantum computing domain. Notably, our classification does not cover fault injection analysis methods, as no such methods have been reported in the literature for quantum~computers.

Furthermore, the fault target and fault manifestation security pyramid for superconducting quantum computers (Figure~\ref{fig_fault_targets}) presented in our work is the quantum computing counterpart of the one introduced by Verbauwhede et al.~\cite{verbauwhede2011fault}.

\section{Conclusion}

This paper presented the first classification of fault-injection attacks on quantum computers. This work first introduced the domain of quantum computer fault injection attacks. It then proceeded to present fault targets and fault manifestations for quantum computers. The resulting classification also specifies fault models unique to quantum computers, along with fault bounds and fault lifespans that should be considered. By shedding light on the vulnerabilities of quantum computers to fault-injection attacks, this work contributes to the development of secure quantum computer systems.

\begin{acks}
The authors would like to thank Yao Lu for suggestions about potential flux trapping faults.
\end{acks}

\bibliographystyle{ACM-Reference-Format}
\bibliography{bibliography}


\begin{thebibliography}{26}


\ifx \showCODEN    \undefined \def \showCODEN     #1{\unskip}     \fi
\ifx \showDOI      \undefined \def \showDOI       #1{#1}\fi
\ifx \showISBNx    \undefined \def \showISBNx     #1{\unskip}     \fi
\ifx \showISBNxiii \undefined \def \showISBNxiii  #1{\unskip}     \fi
\ifx \showISSN     \undefined \def \showISSN      #1{\unskip}     \fi
\ifx \showLCCN     \undefined \def \showLCCN      #1{\unskip}     \fi
\ifx \shownote     \undefined \def \shownote      #1{#1}          \fi
\ifx \showarticletitle \undefined \def \showarticletitle #1{#1}   \fi
\ifx \showURL      \undefined \def \showURL       {\relax}        \fi
\providecommand\bibfield[2]{#2}
\providecommand\bibinfo[2]{#2}
\providecommand\natexlab[1]{#1}
\providecommand\showeprint[2][]{arXiv:#2}

\bibitem[goo(2021)]%
        {google2021exponential}
 \bibinfo{year}{2021}\natexlab{}.
\newblock \showarticletitle{Exponential suppression of bit or phase errors with
  cyclic error correction}.
\newblock \bibinfo{journal}{\emph{Nature}} \bibinfo{volume}{595},
  \bibinfo{number}{7867} (\bibinfo{year}{2021}), \bibinfo{pages}{383--387}.
\newblock


\bibitem[{Amazon Braket SDK}(2023)]%
        {braket_sdk}
\bibfield{author}{\bibinfo{person}{{Amazon Braket SDK}}.}
  \bibinfo{year}{2023}\natexlab{}.
\newblock
\newblock
\newblock
\shownote{\url{https://docs.aws.amazon.com/braket/latest/developerguide/api-and-sdk-reference.html}}.


\bibitem[{Amazon Web Services}(2023)]%
        {amazonbracket}
\bibfield{author}{\bibinfo{person}{{Amazon Web Services}}.}
  \bibinfo{year}{2023}\natexlab{}.
\newblock \bibinfo{title}{{Amazon Braket}}.
\newblock
\newblock
\urldef\tempurl%
\url{https://aws.amazon.com/braket/}
\showURL{%
\tempurl}


\bibitem[Arute et~al\mbox{.}(2019)]%
        {arute2019quantum}
\bibfield{author}{\bibinfo{person}{Frank Arute}, \bibinfo{person}{Kunal Arya},
  \bibinfo{person}{Ryan Babbush}, \bibinfo{person}{Dave Bacon},
  \bibinfo{person}{Joseph~C Bardin}, \bibinfo{person}{Rami Barends},
  \bibinfo{person}{Rupak Biswas}, \bibinfo{person}{Sergio Boixo},
  \bibinfo{person}{Fernando~GSL Brandao}, \bibinfo{person}{David~A Buell},
  {et~al\mbox{.}}} \bibinfo{year}{2019}\natexlab{}.
\newblock \showarticletitle{Quantum supremacy using a programmable
  superconducting processor}.
\newblock \bibinfo{journal}{\emph{Nature}} \bibinfo{volume}{574},
  \bibinfo{number}{7779} (\bibinfo{year}{2019}), \bibinfo{pages}{505--510}.
\newblock


\bibitem[Baksi et~al\mbox{.}(2022)]%
        {baksi2022survey}
\bibfield{author}{\bibinfo{person}{Anubhab Baksi}, \bibinfo{person}{Shivam
  Bhasin}, \bibinfo{person}{Jakub Breier}, \bibinfo{person}{Dirmanto Jap},
  {and} \bibinfo{person}{Dhiman Saha}.} \bibinfo{year}{2022}\natexlab{}.
\newblock \showarticletitle{A survey on fault attacks on symmetric key
  cryptosystems}.
\newblock \bibinfo{journal}{\emph{Comput. Surveys}} \bibinfo{volume}{55},
  \bibinfo{number}{4} (\bibinfo{year}{2022}), \bibinfo{pages}{1--34}.
\newblock


\bibitem[Biamonte et~al\mbox{.}(2017)]%
        {biamonte2017quantum}
\bibfield{author}{\bibinfo{person}{Jacob Biamonte}, \bibinfo{person}{Peter
  Wittek}, \bibinfo{person}{Nicola Pancotti}, \bibinfo{person}{Patrick
  Rebentrost}, \bibinfo{person}{Nathan Wiebe}, {and} \bibinfo{person}{Seth
  Lloyd}.} \bibinfo{year}{2017}\natexlab{}.
\newblock \showarticletitle{Quantum machine learning}.
\newblock \bibinfo{journal}{\emph{Nature}} \bibinfo{volume}{549},
  \bibinfo{number}{7671} (\bibinfo{year}{2017}), \bibinfo{pages}{195--202}.
\newblock


\bibitem[Deutsch et~al\mbox{.}(1995)]%
        {doi:10.1098/rspa.1995.0065}
\bibfield{author}{\bibinfo{person}{David~Elieser Deutsch},
  \bibinfo{person}{Adriano Barenco}, {and} \bibinfo{person}{Artur Ekert}.}
  \bibinfo{year}{1995}\natexlab{}.
\newblock \showarticletitle{Universality in quantum computation}.
\newblock \bibinfo{journal}{\emph{Proceedings of the Royal Society of London.
  Series A: Mathematical and Physical Sciences}} \bibinfo{volume}{449},
  \bibinfo{number}{1937} (\bibinfo{year}{1995}), \bibinfo{pages}{669--677}.
\newblock
\urldef\tempurl%
\url{https://doi.org/10.1098/rspa.1995.0065}
\showDOI{\tempurl}
\showeprint{https://royalsocietypublishing.org/doi/pdf/10.1098/rspa.1995.0065}


\bibitem[Developers(2022)]%
        {cirq_developers_2022}
\bibfield{author}{\bibinfo{person}{Cirq Developers}.}
  \bibinfo{year}{2022}\natexlab{}.
\newblock \showarticletitle{Cirq}.
\newblock  (\bibinfo{date}{Dec} \bibinfo{year}{2022}).
\newblock
\urldef\tempurl%
\url{https://doi.org/10.5281/zenodo.7465577}
\showDOI{\tempurl}
\newblock
\shownote{See full list of authors on Github:
  https://github.com/quantumlib/Cirq/graphs/contributors}.


\bibitem[Devitt et~al\mbox{.}(2013)]%
        {Devitt_2013}
\bibfield{author}{\bibinfo{person}{Simon~J Devitt}, \bibinfo{person}{William~J
  Munro}, {and} \bibinfo{person}{Kae Nemoto}.} \bibinfo{year}{2013}\natexlab{}.
\newblock \showarticletitle{Quantum error correction for beginners}.
\newblock \bibinfo{journal}{\emph{Reports on Progress in Physics}}
  \bibinfo{volume}{76}, \bibinfo{number}{7} (\bibinfo{date}{jun}
  \bibinfo{year}{2013}), \bibinfo{pages}{076001}.
\newblock
\urldef\tempurl%
\url{https://doi.org/10.1088/0034-4885/76/7/076001}
\showDOI{\tempurl}


\bibitem[Giraud and Thiebeauld(2004)]%
        {giraud2004survey}
\bibfield{author}{\bibinfo{person}{Christophe Giraud} {and}
  \bibinfo{person}{Hugues Thiebeauld}.} \bibinfo{year}{2004}\natexlab{}.
\newblock \showarticletitle{A survey on fault attacks}. In
  \bibinfo{booktitle}{\emph{Smart Card Research and Advanced Applications VI:
  IFIP 18th World Computer Congress TC8/WG8. 8 \& TC11/WG11. 2 Sixth
  International Conference on Smart Card Research and Advanced Applications
  (CARDIS) 22--27 August 2004 Toulouse, France}}. Springer,
  \bibinfo{pages}{159--176}.
\newblock


\bibitem[Grover(1996)]%
        {10.1145/237814.237866}
\bibfield{author}{\bibinfo{person}{Lov~K. Grover}.}
  \bibinfo{year}{1996}\natexlab{}.
\newblock \showarticletitle{A Fast Quantum Mechanical Algorithm for Database
  Search}. In \bibinfo{booktitle}{\emph{Proceedings of the Twenty-Eighth Annual
  ACM Symposium on Theory of Computing}} (Philadelphia, Pennsylvania, USA)
  \emph{(\bibinfo{series}{STOC '96})}. \bibinfo{publisher}{Association for
  Computing Machinery}, \bibinfo{address}{New York, NY, USA},
  \bibinfo{pages}{212–219}.
\newblock
\showISBNx{0897917855}
\urldef\tempurl%
\url{https://doi.org/10.1145/237814.237866}
\showDOI{\tempurl}


\bibitem[{IBM}(2022)]%
        {ibmFullPower}
\bibfield{author}{\bibinfo{person}{{IBM}}.} \bibinfo{year}{2022}\natexlab{}.
\newblock \bibinfo{title}{{Bringing the full power of dynamic circuits to
  Qiskit Runtime}}.
\newblock
\newblock
\urldef\tempurl%
\url{https://research.ibm.com/blog/quantum-dynamic-circuits}
\showURL{%
\tempurl}


\bibitem[{IBM}(2023)]%
        {ibmChartingCourse}
\bibfield{author}{\bibinfo{person}{{IBM}}.} \bibinfo{year}{2023}\natexlab{}.
\newblock \bibinfo{title}{{Charting the course to 100,000 qubits}}.
\newblock
\newblock
\urldef\tempurl%
\url{https://research.ibm.com/blog/100k-qubit-supercomputer}
\showURL{%
\tempurl}


\bibitem[{IBM Quantum}(2023)]%
        {ibmquantum}
\bibfield{author}{\bibinfo{person}{{IBM Quantum}}.}
  \bibinfo{year}{2023}\natexlab{}.
\newblock
\newblock
\urldef\tempurl%
\url{https://quantum-computing.ibm.com/}
\showURL{%
\tempurl}


\bibitem[Kim et~al\mbox{.}(2023)]%
        {kim2023evidence}
\bibfield{author}{\bibinfo{person}{Youngseok Kim}, \bibinfo{person}{Andrew
  Eddins}, \bibinfo{person}{Sajant Anand}, \bibinfo{person}{Ken~Xuan Wei},
  \bibinfo{person}{Ewout Van Den~Berg}, \bibinfo{person}{Sami Rosenblatt},
  \bibinfo{person}{Hasan Nayfeh}, \bibinfo{person}{Yantao Wu},
  \bibinfo{person}{Michael Zaletel}, \bibinfo{person}{Kristan Temme},
  {et~al\mbox{.}}} \bibinfo{year}{2023}\natexlab{}.
\newblock \showarticletitle{Evidence for the utility of quantum computing
  before fault tolerance}.
\newblock \bibinfo{journal}{\emph{Nature}} \bibinfo{volume}{618},
  \bibinfo{number}{7965} (\bibinfo{year}{2023}), \bibinfo{pages}{500--505}.
\newblock


\bibitem[{Microsoft Azure}(2023)]%
        {azurequantum}
\bibfield{author}{\bibinfo{person}{{Microsoft Azure}}.}
  \bibinfo{year}{2023}\natexlab{}.
\newblock \bibinfo{title}{{Azure Quantum}}.
\newblock
\newblock
\urldef\tempurl%
\url{https://azure.microsoft.com/en-us/products/quantum}
\showURL{%
\tempurl}


\bibitem[Nielsen and Chuang(2001)]%
        {nielsen2001quantum}
\bibfield{author}{\bibinfo{person}{Michael~A Nielsen} {and}
  \bibinfo{person}{Isaac~L Chuang}.} \bibinfo{year}{2001}\natexlab{}.
\newblock \showarticletitle{Quantum computation and quantum information}.
\newblock \bibinfo{journal}{\emph{Phys. Today}} \bibinfo{volume}{54},
  \bibinfo{number}{2} (\bibinfo{year}{2001}), \bibinfo{pages}{60}.
\newblock


\bibitem[Oliveira et~al\mbox{.}(2022)]%
        {9833778}
\bibfield{author}{\bibinfo{person}{D. Oliveira}, \bibinfo{person}{E. Giusto},
  \bibinfo{person}{E. Dri}, \bibinfo{person}{N. Casciola}, \bibinfo{person}{B.
  Baheri}, \bibinfo{person}{Q. Guan}, \bibinfo{person}{B. Montrucchio}, {and}
  \bibinfo{person}{P. Rech}.} \bibinfo{year}{2022}\natexlab{}.
\newblock \showarticletitle{QuFI: a Quantum Fault Injector to Measure the
  Reliability of Qubits and Quantum Circuits}. In
  \bibinfo{booktitle}{\emph{2022 52nd Annual IEEE/IFIP International Conference
  on Dependable Systems and Networks (DSN)}}. \bibinfo{publisher}{IEEE Computer
  Society}, \bibinfo{address}{Los Alamitos, CA, USA},
  \bibinfo{pages}{137--149}.
\newblock
\urldef\tempurl%
\url{https://doi.org/10.1109/DSN53405.2022.00025}
\showDOI{\tempurl}


\bibitem[Preskill(2018)]%
        {Preskill2018quantumcomputingin}
\bibfield{author}{\bibinfo{person}{John Preskill}.}
  \bibinfo{year}{2018}\natexlab{}.
\newblock \showarticletitle{Quantum {C}omputing in the {NISQ} era and beyond}.
\newblock \bibinfo{journal}{\emph{{Quantum}}}  \bibinfo{volume}{2}
  (\bibinfo{date}{Aug.} \bibinfo{year}{2018}), \bibinfo{pages}{79}.
\newblock
\showISSN{2521-327X}
\urldef\tempurl%
\url{https://doi.org/10.22331/q-2018-08-06-79}
\showDOI{\tempurl}


\bibitem[{Qiskit contributors}(2023)]%
        {Qiskit}
\bibfield{author}{\bibinfo{person}{{Qiskit contributors}}.}
  \bibinfo{year}{2023}\natexlab{}.
\newblock \bibinfo{title}{Qiskit: An Open-source Framework for Quantum
  Computing}.
\newblock
\newblock
\urldef\tempurl%
\url{https://doi.org/10.5281/zenodo.2573505}
\showDOI{\tempurl}


\bibitem[Ravi et~al\mbox{.}(2022)]%
        {ravi2022side}
\bibfield{author}{\bibinfo{person}{Prasanna Ravi}, \bibinfo{person}{Anupam
  Chattopadhyay}, \bibinfo{person}{Jan~Pieter D'Anvers}, {and}
  \bibinfo{person}{Anubhab Baksi}.} \bibinfo{year}{2022}\natexlab{}.
\newblock \showarticletitle{Side-channel and fault-injection attacks over
  lattice-based post-quantum schemes (Kyber, Dilithium): Survey and new
  results}.
\newblock \bibinfo{journal}{\emph{ACM Transactions on Embedded Computing
  Systems}} (\bibinfo{year}{2022}).
\newblock


\bibitem[Rivest et~al\mbox{.}(1978)]%
        {10.1145/359340.359342}
\bibfield{author}{\bibinfo{person}{R.~L. Rivest}, \bibinfo{person}{A. Shamir},
  {and} \bibinfo{person}{L. Adleman}.} \bibinfo{year}{1978}\natexlab{}.
\newblock \showarticletitle{A Method for Obtaining Digital Signatures and
  Public-Key Cryptosystems}.
\newblock \bibinfo{journal}{\emph{Commun. ACM}} \bibinfo{volume}{21},
  \bibinfo{number}{2} (\bibinfo{date}{feb} \bibinfo{year}{1978}),
  \bibinfo{pages}{120–126}.
\newblock
\showISSN{0001-0782}
\urldef\tempurl%
\url{https://doi.org/10.1145/359340.359342}
\showDOI{\tempurl}


\bibitem[Shepherd et~al\mbox{.}(2021)]%
        {shepherd2021physical}
\bibfield{author}{\bibinfo{person}{Carlton Shepherd},
  \bibinfo{person}{Konstantinos Markantonakis}, \bibinfo{person}{Nico van
  Heijningen}, \bibinfo{person}{Driss Aboulkassimi},
  \bibinfo{person}{Cl{\'e}ment Gaine}, \bibinfo{person}{Thibaut Heckmann},
  {and} \bibinfo{person}{David Naccache}.} \bibinfo{year}{2021}\natexlab{}.
\newblock \showarticletitle{Physical fault injection and side-channel attacks
  on mobile devices: A comprehensive analysis}.
\newblock \bibinfo{journal}{\emph{Computers \& Security}}
  \bibinfo{volume}{111} (\bibinfo{year}{2021}), \bibinfo{pages}{102471}.
\newblock


\bibitem[Shor(1997)]%
        {doi:10.1137/S0097539795293172}
\bibfield{author}{\bibinfo{person}{Peter~W. Shor}.}
  \bibinfo{year}{1997}\natexlab{}.
\newblock \showarticletitle{Polynomial-Time Algorithms for Prime Factorization
  and Discrete Logarithms on a Quantum Computer}.
\newblock \bibinfo{journal}{\emph{SIAM J. Comput.}} \bibinfo{volume}{26},
  \bibinfo{number}{5} (\bibinfo{year}{1997}), \bibinfo{pages}{1484--1509}.
\newblock
\urldef\tempurl%
\url{https://doi.org/10.1137/S0097539795293172}
\showDOI{\tempurl}
\showeprint{https://doi.org/10.1137/S0097539795293172}


\bibitem[Veps{\"a}l{\"a}inen et~al\mbox{.}(2020)]%
        {vepsalainen2020impact}
\bibfield{author}{\bibinfo{person}{Antti~P Veps{\"a}l{\"a}inen},
  \bibinfo{person}{Amir~H Karamlou}, \bibinfo{person}{John~L Orrell},
  \bibinfo{person}{Akshunna~S Dogra}, \bibinfo{person}{Ben Loer},
  \bibinfo{person}{Francisca Vasconcelos}, \bibinfo{person}{David~K Kim},
  \bibinfo{person}{Alexander~J Melville}, \bibinfo{person}{Bethany~M
  Niedzielski}, \bibinfo{person}{Jonilyn~L Yoder}, {et~al\mbox{.}}}
  \bibinfo{year}{2020}\natexlab{}.
\newblock \showarticletitle{Impact of ionizing radiation on superconducting
  qubit coherence}.
\newblock \bibinfo{journal}{\emph{Nature}} \bibinfo{volume}{584},
  \bibinfo{number}{7822} (\bibinfo{year}{2020}), \bibinfo{pages}{551--556}.
\newblock


\bibitem[Verbauwhede et~al\mbox{.}(2011)]%
        {verbauwhede2011fault}
\bibfield{author}{\bibinfo{person}{Ingrid Verbauwhede}, \bibinfo{person}{Dusko
  Karaklajic}, {and} \bibinfo{person}{Jorn-Marc Schmidt}.}
  \bibinfo{year}{2011}\natexlab{}.
\newblock \showarticletitle{The fault attack jungle-a classification model to
  guide you}. In \bibinfo{booktitle}{\emph{2011 Workshop on Fault Diagnosis and
  Tolerance in Cryptography}}. IEEE, \bibinfo{pages}{3--8}.
\newblock


\end{thebibliography}

\end{document}